\documentclass[twocolumn,showpacs,preprintnumbers,amsmath,amssymb,pre]{revtex4}

\usepackage{graphicx}
\usepackage{dcolumn}
\usepackage{bm}
\usepackage{enumerate}

\begin{document}

\title{Subdiffusion--absorption process in a system consisting of two different media}

\author{Tadeusz Koszto{\l}owicz}
 \email{tadeusz.kosztolowicz@ujk.edu.pl}
 \affiliation{Institute of Physics, Jan Kochanowski University,\\
         ul. \'Swi\c{e}tokrzyska 15, 25-406 Kielce, Poland}

\date{\today}

\begin{abstract}
Subdiffusion with reaction $A+B\rightarrow B$ is considered in a system which consists of two homogeneous media joined together; the $A$ particles are mobile whereas $B$ are static. Subdiffusion and reaction parameters, which are assumed to be independent of time and space variable, can be different in both media. Particles $A$ move freely across the border between the media. In each part of the system the process is described by the subdiffusion--reaction equations with fractional time derivative. By means of the method presented in this paper we derive both the fundamental solutions (the Green's functions) $P(x,t)$ to the subdiffusion--reaction equations and the boundary conditions at the border between the media. One of the conditions demands the continuity of a flux and the other one contains the Riemann--Liouville fractional time derivatives $\partial^{\alpha_1}P(0^+,t)/\partial t^{\alpha_1}=(D_1/D_2)\partial^{\alpha_2}P(0^-,t)/\partial t^{\alpha_2}$, where the subdiffusion parameters $\alpha_1$, $D_1$ and $\alpha_2$, $D_2$ are defined in the regions $x<0$ and $x>0$, respectively.
\end{abstract}


\maketitle

\section{Introduction\label{secI}}

Subdiffusion can occur in media, as gels or porous media, in which random walk of particles is significantly hindered by a complex structure of a medium \cite{mk,kdm}. In a three dimensional system subdiffusion is often characterized by the relation $\left\langle (\Delta \vec{r})^2\right\rangle=6Dt^\alpha/\Gamma(1+\alpha)$, where $\left\langle (\Delta \vec{r})^2\right\rangle$ denotes the mean square displacement of a particle, $D$ is a subdiffusion coefficient and $\alpha$ is a subdiffusion parameter, $\Gamma$ denotes the Gamma function, $0<\alpha<1$. The case $\alpha=1$ corresponds to normal diffusion. We consider subdiffusion with reaction in a system which consists of two homogeneous media joined together. 
We assume that static particles $B_1$ are homogeneously distributed in a first medium whereas static particles $B_2$ are homogeneously distributed in a second medium. A mobile particle $A$ can react with particles $B_1$ and $B_2$ according to the formula $A+B_i\rightarrow B_i$, $i=1,2$.
The reactions can be interpreted as absorption of particles $A$ by static `absorbing centres' $B_1$ and $B_2$. We mention here that normal diffusion and anomalous diffusion with absorption have been studied in zeolites \cite{smit} and in clay materials such as bentonite \cite{bourg} and halloysite \cite{luo}. The absorbing centers may be located on the surfaces of tubules inside the medium in which the diffusion takes place. We add that, as it is discussed in \cite{guimaraes}, presence of absorbing centers on walls bounding the system can change a character of diffusion. Subdiffusion with reactions (absorption) is usually described by equations with fractional time derivative \cite{sung,seki,ks,mendez,yl}. When a particle $A$ meets $B_i$ the reaction can occur with probability controlled by reaction rates: $\kappa^2_1$ in the first medium and $\kappa^2_2$ in the second one. 

The situation is complicated when subdiffusion is considered in a composite system which consists of different media joined together. To solve equations describing subdiffusion--reaction process in a composite medium one needs two boundary conditions at the border between media. However, there are many examples that the determination of the boundary conditions at the points of system's discontinuity is ambiguous \cite{boundary,korabel}.

\begin{figure}[h!]
\centering
\includegraphics[scale=0.5]{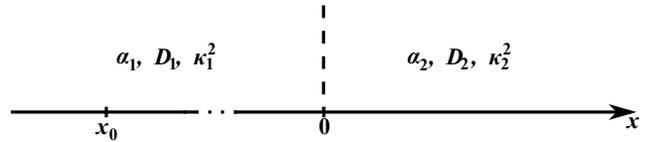}
\caption{The system under consideration. There are various subdiffusion parameters and reaction rates in different media, the plane $x=0$ is the border between the media, $x_0$ is the initial position of the particle. \label{r1}}
\end{figure}

In this paper we consider subdiffusion in a composite system consisting of two media which can have different subdiffusion parameters and reaction rates, see Fig. \ref{r1}. 
Our considerations concern a three--dimensional system which is homogeneous in the plane perpendicular to the $x$-axis. Thus, later in this paper we treat this system as effectively one--dimensional.

We suppose that equations describing the process are the following \cite{seki}
\begin{eqnarray}\label{eq1}
\frac{\partial P_-(x,t;x_0)}{\partial t}= 
D_1\frac{\partial^{1-\alpha_1}}{\partial t^{1-\alpha_1}}\Bigg[\frac{\partial^2 P_-(x,t;x_0)}{\partial x^2}\\
-\kappa_1^2 P_-(x,t;x_0)\Bigg],\nonumber
\end{eqnarray}

\begin{eqnarray}\label{eq2}
\frac{\partial P_+(x,t;x_0)}{\partial t}= 
D_2\frac{\partial^{1-\alpha_2}}{\partial t^{1-\alpha_2}}\Bigg[\frac{\partial^2 P_+(x,t;x_0)}{\partial x^2}\\
-\kappa_2^2 P_+(x,t;x_0)\Bigg],\nonumber
\end{eqnarray}
where $P_\pm(x,t;x_0)$ denotes a probability density of finding the particle at point $x$ at time $t$, $x_0$ is the initial position of the particle and the indexes $-$ and $+$ mean that the current position of the particle $x$ is located at the regions $x<0$ and $x>0$, respectively. 
The probability fulfills the initial condition $P_\pm(x,0;x_0)=\delta(x-x_0)$, where $\delta$ is the Dirac--delta function; in the following we assume $x_0<0$.
The Riemann--Liouville derivative, occurring in the above equations, is defined as being valid for $\delta>0$ (here $k$ is a natural number which fulfils $k-1\leq \delta <k$)
\begin{equation}\label{eq3}
\frac{d^\delta}{dt^\delta}f(t)=\frac{1}{\Gamma(k-\delta)}\frac{d^k}{dt^k}\int_0^t{(t-t')^{k-\delta-1}f(t')dt'}\;.
\end{equation}

We assume that the particles $A$ can pass freely across the border between the media. It means that a particle which makes a jump from one medium to another through the boundary between the media will come to a new medium. There are no obstacles such as partially permeable barrier between the media. An `anomalous behavior' of the particle in the border region can be created only by the difference of subdiffusion and reaction parameters of the media.

The main aim is to derive the Green's functions for the system under study. From the obtained functions we derive boundary conditions at the border between media. 
In order to derive the Green's functions we will use a simple model of a random walk with reactions in a system with discrete time and spatial variables. 
Next, we move to a system with continuous variables by means of the rules presented in this paper. The choice of such methodology is due to the fact that difference equations describing random walk of particle in a composite system are solvable. The behavior of the particle at the boundary between media is involved in discrete model in a `naturaly way'.
Similar models were previously used to derive the Green's functions for various kinds of subdiffusion in a system with a thin membrane \cite{tk1,tk3} and to derive parabolic or hiperbolic subdiffusion--reaction equations \cite{kl,tk4}. However, the procedure used here is significantly changed compared to the procedures used in the above cited papers. The reason is that when considering a system consisting of two different media, additional rules which enable to correctly set the reaction and subdiffusion parameters of both media in the obtained functions should be found.

The organization of the paper is as follows. In Sec. \ref{secII} we present the general procedure which is used in subsequent considerations. In this section we consider subdiffusion--reaction process in a homogeneous system. We compare the difference equation, which describes random walk with absorption in a system with discrete time and space variables, and the subdiffusion--reaction equation with fractional time derivative, which describes subdiffusion with absorption in a system with continuous variables. In particular, there will be shown the relations between diffusion and absorption parameters defined in both systems. This section does not have new results but within it we present some details of the procedure. The random walk model with reactions in a discrete system consisting of two media is considered in Sec. \ref{secIII}. The Laplace transforms of Green's functions and boundary conditions at the border between media are derived in Sec. \ref{secIV}. In this section we also show an example of calculating the Green's functions in time domain from their Laplace transforms. The final remarks are presented in Sec. \ref{secV}.

\section{Method\label{secII}}

We start our consideration with difference equations which describe random walk in a system with discrete time and space variable. In this section we discuss the relation between subdiffusion--reaction equations defined in systems with continuous variables and discrete variables. 
The subdiffusion--reaction equation which describes the process in a homogeneous system with continuous variables reads
\begin{eqnarray}\label{eq4}
\frac{\partial P(x,t;x_0)}{\partial t}= 
D\frac{\partial^{1-\alpha}}{\partial t^{1-\alpha}}\Bigg[\frac{\partial^2 P(x,t;x_0)}{\partial x^2}\\
-\kappa^2 P(x,t;x_0)\Bigg].\nonumber
\end{eqnarray}
The Laplace transform of the fundamental solution to Eq. (\ref{eq4}) is
\begin{equation}\label{eq5}
\hat{P}(x,s;x_0)=\frac{s^{\alpha-1}}{2D\sqrt{\kappa^2+\frac{s^\alpha}{D}}}{\rm e}^{-\sqrt{\kappa^2+\frac{s^\alpha}{D}}|x-x_0|}.
\end{equation}

The function (\ref{eq5}) can be derived from a simple model which describes particle's random walk with reaction in a system with discrete variables. The model is based on the difference equation 
\begin{eqnarray}\label{eq6}
P_{n+1}(m;m_0)=\frac{1}{2}P_n(m+1;m_0)+\frac{1}{2}P_n(m-1;m_0)
\\-R P_n(m;m_0)\;,\nonumber
\end{eqnarray}
where $P_n(m;m_0)$ is a probability of finding the particle in a site $m$ after $n$ steps, $m_0$ is the initial position of the particle, $R$ is a the probability of absorption of particle during its stopover at a current position. The initial condition reads $P_0(m;m_0)=\delta_{m,m_0}$ where $\delta$ denotes here the Kronecker symbol.
The generating function is defined as $S(z,m;m_0)=\sum_{n=0}^\infty z^n P_n(m;m_0)$.
Passing from discrete to continuous time we use the formula \cite{mk,ks,mw} $P(m,t;m_0)=\int_0^t U(t-t')\sum_{n=0}^\infty Q_n(t')P_n(m;m_0)dt'$, where $Q_n(t')$ is the probability that a particle takes $n$ steps in the time interval $(0,t')$ and the last step is performed exactly in time $t'$, $Q_{n+1}(t)=\int_0^t \omega(t-t')Q_n(t')dt'$ for $n\geq 1$, $Q_0(t)=\delta(t)$, $\omega(t)$ is a distribution of time which is needed to take particle's next step,  $U(t)=1-\int_0^{t}\omega(t')dt'$ denotes the probability that the paricle stays in its current position up to time $t$. In terms of the Laplace transform, $\mathcal{L}[f(t)]\equiv\hat{f}(s)=\int_0^\infty {\rm exp}(-st)f(t)dt$, we have $\hat{Q}_n(s)=\hat{\omega}^n(s)$ \cite{mk} and 
\begin{equation}\label{eq7}
U(s)=\frac{1-\hat{\omega}(s)}{s}. 
\end{equation}
Combining the above formulas one gets the well-known formula 
\begin{equation}\label{eq8}
\hat{P}(m;s;m_0)=\hat{U}(s)S(m,\hat{\omega}(s);m_0).
\end{equation} 
To pass from discrete $m$ to continuous $x$ position one assumes that $x=\epsilon m$ and $x_0=\epsilon m_0$, where the distance between discrete sites $\epsilon$ is assumed to be small but nonzero. The probability density of finding a particle at point $x$ at time $t$ is 
\begin{equation}\label{eq9}
P(x,t;x_0)=\frac{P(m,t;m_0)}{\epsilon}. 
\end{equation}
The generating function of Eq. (\ref{eq6}) reads \cite{kl}
\begin{equation}\label{eq10}
S(m,z;m_0)=\frac{[\eta_R(z)]^{|m-m_0|}}{\sqrt{(1+zR)^2-z^2}}\;,
\end {equation}
where
\begin{equation}\label{eq11}
\eta_R(z)=\frac{1+zR-\sqrt{(1+zR)^2-z^2}}{z}\;.
\end{equation}
Usually it is assumed that \cite{mk}
\begin{equation}\label{eq12}
\hat{\omega}(s)=1-\tau s^\alpha,
\end{equation} 
where $0<\alpha<1$ for subdiffusion and $\alpha=1$ for normal diffussion, $\tau$ is a positive parameter. The form of the function $\hat{\omega}$, Eq. (\ref{eq12}), is motivated by the fact that consideration is usually conducted in a limit of long time which corresponds to the limit of the small parameter $s$. However, a timescale for subdiffusion is not defined clearly. Since the subdiffusion coefficient is defined as $D=\epsilon^2/(2\tau)$ \cite{tk1,kl}, Eq. (\ref{eq12}) can be interpreted as an approximation of $\hat{\omega}$ in the limit of small $\epsilon$ (or, equivalently, in the limit of small parameter $\tau$) valid for $s>0$. From the above equations we get $\eta_R(\hat{\omega}(s))\approx 1+R+\epsilon^2 s^\alpha/(2D)-\sqrt{(1+R+\epsilon^2 s^\alpha/(2D))^2-1}$. 
We obtain the function (\ref{eq5}) from Eqs. (\ref{eq6})-(\ref{eq11}) and the above equations in the limit of small $\epsilon$ only if 
\begin{equation}\label{eq13}
R=\frac{\epsilon^2\kappa^2}{2} .
\end{equation} 
Eq. (\ref{eq13}) links the reaction parameters defined in the systems with discrete and continuous spatial variable. This result was also derived in \cite{kl} using different arguments than presented in this paper.

\section{Model\label{secIII}}

\begin{figure}[h!]
\centering
\includegraphics[scale=0.45]{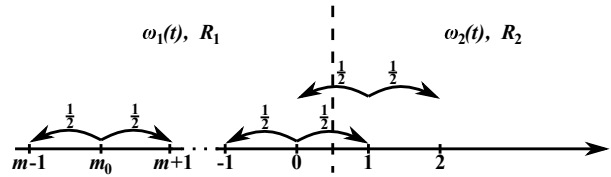}
\caption{Random walk in a discrete system which consists of two media, the border between media is represented by the verical dashed line, the functions $\omega_1$ and $\omega_2$ are the probability densities of waiting time to take particle's next step in appropariate parts of the system. \label{r2}}
\end{figure}

Random walk in a system with discrete both time and space variables can be described by the following difference equations (see Fig. \ref{r2})

 \begin{eqnarray}\label{eq14}
P^-_{n+1}(m;m_0)=\frac{1}{2}P^-_n(m+1;m_0)+\frac{1}{2}P^-_n(m-1;m_0)\nonumber\\
-R_1 P^-_n(m;m_0),\;m\leq -1\;,
	\end{eqnarray}	
 \begin{eqnarray}\label{eq15}
P^-_{n+1}(0;m_0)=\frac{1}{2}P^+_n(1;m_0)+\frac{1}{2}P^-_n(-1;m_0)\\-R_1 P^-_n(0;m_0)\;,\nonumber
	\end{eqnarray}
\begin{eqnarray}\label{eq16}
P^+_{n+1}(1;m_0)=\frac{1}{2}P^+_n(2;m_0)+\frac{1}{2}P^-_n(0;m_0)\\-R_2 P^+_n(1;m_0)\;,\nonumber
\end{eqnarray}
\begin{eqnarray}\label{eq17}
P^+_{n+1}(m;m_0)=\frac{1}{2}P^+_n(m+1;m_0)+\frac{1}{2}P^+_n(m-1;m_0)\nonumber\\
-R_2 P^+_n(m;m_0),\;m\geq 2\;,
\end{eqnarray}
where $R_1$ and $R_2$ denote the probability of partcle's absorption during its stay at a site located in the left--hand side and the right--hand side of the system, respectively. In order to pass to continuous time we use the generating function $S_{\pm}(m,z;m_0)=\sum_{n=0}^\infty z^n P^{\pm}_n(m,z;m_0)$.
After calculation we get (the details of calculation are presented in \cite{tk3})
\begin{eqnarray}\label{eq18}
S_-(m,z;m_0)=\frac{[\eta_{R_1}(z)]^{|m-m_0|}}{\sqrt{(1+zR_1)^2-z^2}}\\
+\eta_{R_1}(z)\left(\frac{\eta_{R_2}(z)-\eta_{R_1}(z)}{1-\eta_{R_1}(z)\eta_{R_2}(z)}\right)\frac{[\eta_{R_1}(z)]^{2N-m-m_0}}{\sqrt{(1+zR_1)^2-z^2}},\nonumber
\end{eqnarray}
\begin{eqnarray}\label{eq19}
S_+(m,z;m_0)=\frac{[\eta_{R_1}(z)]^{N-m_0}[\eta_{R_2}(z)]^{m-N-1}}{\sqrt{(1+zR_2)^2-z^2}}\\
\times\eta_{R_1}(z)\left(\frac{1-\eta^2_{R_1}(z)}{1-\eta_{R_1}(z)\eta_{R_2}(z)}\right).\nonumber
\end{eqnarray}

In homogeneous system in order to pass to continuous time we make the conversion $z\rightarrow \hat{\omega}(s)$ in the generating function. However, the situation is more complicated in a system which consists of two parts with different transport properties. The reason is that $\hat{\omega}_1(s)$ and $\hat{\omega}_2(s)$ should be involved into the generating functions. As in Sec. \ref{secII}, we suppose
\begin{eqnarray}\label{eq20}
\hat{P}_-(m,s;m_0)=\hat{U}_1(s)S_-(m,\{\hat{\omega}_1(s),\hat{\omega}_2(s)\};m_0),
\end{eqnarray}
\begin{eqnarray}\label{eq21}
\hat{P}_+(m,s;m_0)=\hat{U}_2(s)S_+(m,\{\hat{\omega}_1(s),\hat{\omega}_2(s)\};m_0),
\end{eqnarray}
where
\begin{equation}\label{eq22}
\hat{U}_i(s)=\frac{1-\hat{\omega}_i(s)}{s},
\end{equation}
$i=1,2$.

In order to find a rule which allows us to incorporate correctly $\hat{\omega}_1$ and $\hat{\omega}_2$ into $S_\pm$, we employ a first passage time distribution. 
The particle which starts from $m_0$ achieves the position $m$ first time after $n$ steps with probability $F_n(m;m_0)$. The generating function $S_F$ for this probability reads
\begin{eqnarray}\label{eq23}
S_F(m,z;m_0)\equiv\sum_{n=1}^\infty z^n F_n(m;m_0)\\
=\frac{S(m,z;m_0)-\delta_{m,m_0}}{S(m,z;m)}.\nonumber
\end{eqnarray}
We add that the equation for generating function $S_F$ derived in \cite{mw}, which has different denominator comparing to Eq. (\ref{eq23}), is valid for the random walk in a homogeneous system only. Supposing $m=-1$ and $m_0<-1$, from Eqs. (\ref{eq18}) and (\ref{eq23}) we get
\begin{equation}\label{eq24}
S_F(-1,z;m_0)=[\eta_{R_1}(z)]^{-1-m_0}.
\end{equation}
All particle's steps performed from $m_0$ to $m=-1$ are ruled by the function $\omega_1$. Moving to continuous time we have $\hat{F}(-1,s;m_0)=\sum_{n=1}^\infty \hat{\omega}_1^n(s) F_n(-1;m_0)\equiv S_F(-1,\hat{\omega}_1(s);m_0)$. Finally, we get
\begin{equation}\label{eq25}
\hat{F}(-1,s;m_0)=[\eta_{R_1}(\hat{\omega}_1(s))]^{-1-m_0}.
\end{equation}
From Eq. (\ref{eq25}) we deduce that the function $\eta_{R_1}$ depends on $\hat{\omega}_1$ only. Due to the symmetry argument, the function $\eta_{R_2}$ depends on $\hat{\omega}_2$ only. Thus, the replacement of $z$ by $\hat{\omega}_1$ and $\hat{\omega}_2$ in Eqs. (\ref{eq18})--(\ref{eq21}) should be performed by means of the following rule
\begin{equation}\label{eq26}
\eta_{R_1}(z)\rightarrow \eta_{R_1}(\hat{\omega}_1(s))\;,\;
\eta_{R_2}(z)\rightarrow \eta_{R_2}(\hat{\omega}_2(s)).
\end{equation}
We suppose $R_i=\epsilon^2 \kappa^2_i/2$, $\hat{\omega}_i(s)=1-\tau_i s^{\alpha_i}$, and $D_i=\epsilon^2/(2\tau_i)$, $i=1,2$. Then, we have
\begin{equation}\label{eq27} 
\hat{\omega}_i(s)=1-\frac{\epsilon^2 s^{\alpha_i}}{2D_i}.
\end{equation} 

\section{Results\label{secIV}}

From Eqs. (\ref{eq9}), (\ref{eq18})--(\ref{eq22}), (\ref{eq26}), and (\ref{eq27}) we get in the limit of small $\epsilon$
\begin{eqnarray}\label{eq28}
\hat{P}_-(x,s;x_0)=\frac{s^{-1+\alpha_1}}{2D_1\sqrt{\kappa^2_1+\frac{s^{\alpha_1}}{D_1}}}\left[{\rm e}^{-\sqrt{\kappa^2_1+\frac{s^{\alpha_1}}{D_1}}|x-x_0|}\right.\\
\left.+(1-\Lambda(s)){\rm e}^{\sqrt{\kappa^2_1+\frac{s^{\alpha_1}}{D_1}}(x+x_0)}\right],\nonumber
\end{eqnarray}
\begin{eqnarray}\label{eq29}
\hat{P}_+(x,s;x_0)=\Lambda(s)\frac{s^{-1+\alpha_2}}{2D_2\sqrt{\kappa^2_2+\frac{s^{\alpha_2}}{D_2}}}\\
\times\left[{\rm e}^{\sqrt{\kappa^2_1+\frac{s^{\alpha_1}}{D_1}}x_0}{\rm e}^{-\sqrt{\kappa^2_2+\frac{s^{\alpha_2}}{D_2}}x}\right],\nonumber
\end{eqnarray}
where
\begin{equation}\label{eq30}
\Lambda(s)=\frac{2\sqrt{\kappa^2_2+\frac{s^{\alpha_2}}{D_2}}}{\sqrt{\kappa^2_1+\frac{s^{\alpha_1}}{D_1}}+\sqrt{\kappa^2_2+\frac{s^{\alpha_2}}{D_2}}}.
\end{equation}
The functions (\ref{eq28}) and (\ref{eq29}) fulfill the following boundary conditions
\begin{equation}\label{eq31}
s^{\alpha_1}\hat{P}_+(0^+,s;x_0)=s^{\alpha_2}\frac{D_1}{D_2}\hat{P}_-(0^-,s;x_0),
\end{equation}
\begin{equation}\label{eq32}
\hat{J}_+(0^+,s;x_0)=\hat{J}_-(0^-,s;x_0),
\end{equation}
where $J$ denotes the subdiffusive flux, the Laplace transforms of fulxes are
\begin{eqnarray}
\hat{J}_-(x,s;x_0)=-D_1 s^{1-\alpha_1}\frac{\partial\hat{P}_-(x,s;x_0)}{\partial x},\label{eq33}\\
\hat{J}_+(x,s;x_0)=-D_2 s^{1-\alpha_2}\frac{\partial\hat{P}_+(x,s;x_0)}{\partial x}.\label{eq34}
\end{eqnarray}
The Laplace transform of the Riemann-Liouville fractional derivative reads \cite{oldham,podlubny}
\begin{equation}\label{eq35}
\mathcal{L}\left[\frac{d^\delta}{d t^\delta}f(t)\right]=s^\delta\hat{f}(s)-\sum_{i=0}^{k-1}s^i f^{(\delta-i-1)}(0)\;,
\end{equation}
where $f^{(\delta-i-1)}(0)$ is the initial value of the derivative of $(\delta-i-1)$--th order. Since this value is often considered as unknown, the relation (\ref{eq35}) is somewhat useless. However, for $0<\delta<1$ and for the case of a bounded function $f$ there is $f^{(\delta)}(0)=0$. Therefore, (\ref{eq35}) reads for this case
\begin{equation}\label{eq36}
\mathcal{L}\left[\frac{d^\delta}{d t^\delta}f(t)\right]=s^\delta\hat{f}(s)\;.
\end{equation}
Eqs. (\ref{eq31})--(\ref{eq36}) provide the following boundary conditions
\begin{equation}\label{eq37}
\frac{\partial^{\alpha_1}P_+(0^+,t;x_0)}{\partial t^{\alpha_1}}=\frac{D_1}{D_2}\frac{\partial^{\alpha_2}P_-(0^-,t;x_0)}{\partial t^{\alpha_2}},
\end{equation}
\begin{equation}\label{eq38}
J_+(0^+,t;x_0)=J_-(0^-,t;x_0),
\end{equation}
where $J_+=-D_2\partial^{1-\alpha_2}P_+/\partial t^{1-\alpha_2}$ and $J_-=-D_1\partial^{1-\alpha_1}P_-/\partial t^{1-\alpha_1}$. The second boundary condition, Eq. (\ref{eq38}), shows that the subdiffusive flux flowing through the border between the media is continuous. However, the first one, Eq. (\ref{eq37}), takes rather unexpected form since it is independent of the reaction parameters. For the case of $\alpha_1=\alpha_2=\alpha$ it appears to be also independent of subdiffusion parameter $\alpha$ (see Eq. (\ref{eq31})) and reads
\begin{equation}\label{eq39}
P_+(0^+,t;x_0)=\frac{D_1}{D_2}P_-(0^-,t;x_0).
\end{equation}

To calculate the inversion Laplace transforms of the Green's functions, Eqs. (\ref{eq28}) and (\ref{eq29}), we use the approximation of small parameter $s$. In the example presented below the approximate functions include the leading terms with respect to $s$ in such a way that all parameters describing the system are contained in the obtained functions. In the calculation we use the following formulas $\sqrt{\kappa^2 +s^\alpha/D}\approx \kappa+s^\alpha/(2D\kappa)$, ${\rm e}^{u}=\sum_{n=0}^\infty u^n/n!$, and \cite{tk}
\begin{eqnarray}\label{eq40}
\mathcal{L}^{-1}\left[s^\nu {\rm e}^{-as^\beta}\right]\equiv f_{\nu,\beta}(t;a)\\
=\frac{1}{t^{\nu+1}}\sum_{k=0}^\infty{\frac{1}{k!\Gamma(-k\beta-\nu)}\left(-\frac{a}{t^\beta}\right)^k},\nonumber
\end{eqnarray}
$a,\beta>0$, the function $f_{\nu,\beta}$ is the special case of the H--Fox function. 
Detailed form of the approximate function depends on the relation between parameters $\alpha_1$ and $\alpha_2$. For $\alpha_2<2\alpha_1$ we get

\begin{eqnarray}\label{eq41}
\lefteqn{P_-(x,t;x_0)=\frac{1}{2D_1\kappa_1}\Bigg[{\rm e}^{-|x-x_0|\kappa_1}f_{\alpha_1-1,\alpha_1}\left(t;\frac{|x-x_0|}{2D_1\kappa_1}\right)}\nonumber\\
& & +\frac{\kappa_1-\kappa_2}{\kappa_1+\kappa_2}{\rm e}^{(x+x_0)\kappa_1}f_{\alpha_1-1,\alpha_1}\left(t;\frac{-(x+x_0)}{2D_1\kappa_1}\right)\Bigg]\\ 
& & \mbox{} -\frac{1}{4D_1^2\kappa_1^3}\Bigg[{\rm e}^{-|x-x_0|\kappa_1}f_{2\alpha_1-1,\alpha_1}\left(t;\frac{|x-x_0|}{2D_1\kappa_1}\right)\nonumber\\
& & +\frac{\kappa_1-\kappa_2}{\kappa_1+\kappa_2}{\rm e}^{(x+x_0)\kappa_1}f_{2\alpha_1-1,\alpha_1}\left(t;\frac{-(x+x_0)}{2D_1\kappa_1}\right)\Bigg]\nonumber\\ 
& & +\frac{{\rm e}^{(x+x_0)\kappa_1}}{2(\kappa_1+\kappa_2)^2 D_1\kappa_1}\Bigg[\frac{\kappa_2}{D_1\kappa_1}f_{2\alpha_1-1,\alpha_1}\left(t;\frac{-(x+x_0)}{2D_1\kappa_1}\right)\nonumber\\
& & -\frac{\kappa_1}{D_2\kappa_2}f_{\alpha_1+\alpha_2-1,\alpha_1}\left(t;\frac{-(x+x_0)}{2D_1\kappa_1}\right)\Bigg],\nonumber
\end{eqnarray}

\begin{eqnarray}\label{eq42}
\lefteqn{P_+(x,t;x_0)=\frac{{\rm e}^{x_0\kappa_1-x\kappa_2}}{D_2(\kappa_1+\kappa_2)}\times}\\
& & \sum_{n=0}^\infty \frac{1}{n!}\left(\frac{x_0}{2D_1\kappa_1}\right)^n\Bigg[f_{n\alpha_1+\alpha_2-1,\alpha_2}\left(t;\frac{x}{2D_2\kappa_2}\right)\nonumber\\
& & \mbox{} -\frac{1}{2D_1\kappa_1(\kappa_1+\kappa_2)}f_{(n+1)\alpha_1+\alpha_2-1,\alpha_2}\left(t;\frac{x}{2D_2\kappa_2}\right)\nonumber\\
& & -\frac{1}{2D_2\kappa_2(\kappa_1+\kappa_2)}f_{n\alpha_1+2\alpha_2-1,\alpha_2}\left(t;\frac{x}{2D_2\kappa_2}\right)\Bigg].\nonumber
\end{eqnarray}

Below there are shown plots of Green's functions Eqs. (\ref{eq41}) and (\ref{eq42}). For all cases $x_0=-1$, $D_1=0.1$, $D_2=0.2$, $\kappa_1=2$, and $\kappa_2=1$, all quantities are given in arbitrary chosen units. 

\begin{figure}[!ht]
\centering
\includegraphics[height=5.7cm]{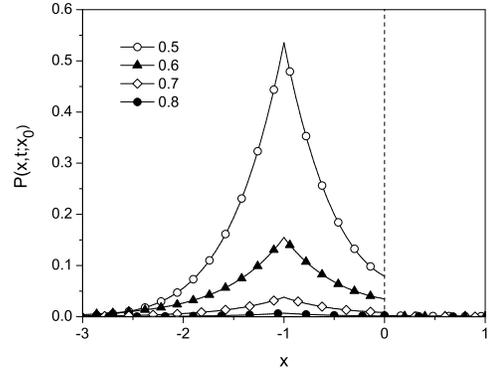}
\caption{The plots of functions (\ref{eq41}) and (\ref{eq42}) for $t=10^4$, $\alpha_2=0.9$, and for various parameter $\alpha_1$ given in the legend. The vertical dashed line represents the border between media.}\label{Fig1}
\end{figure}

\begin{figure}[!ht]
\centering
\includegraphics[height=5.7cm]{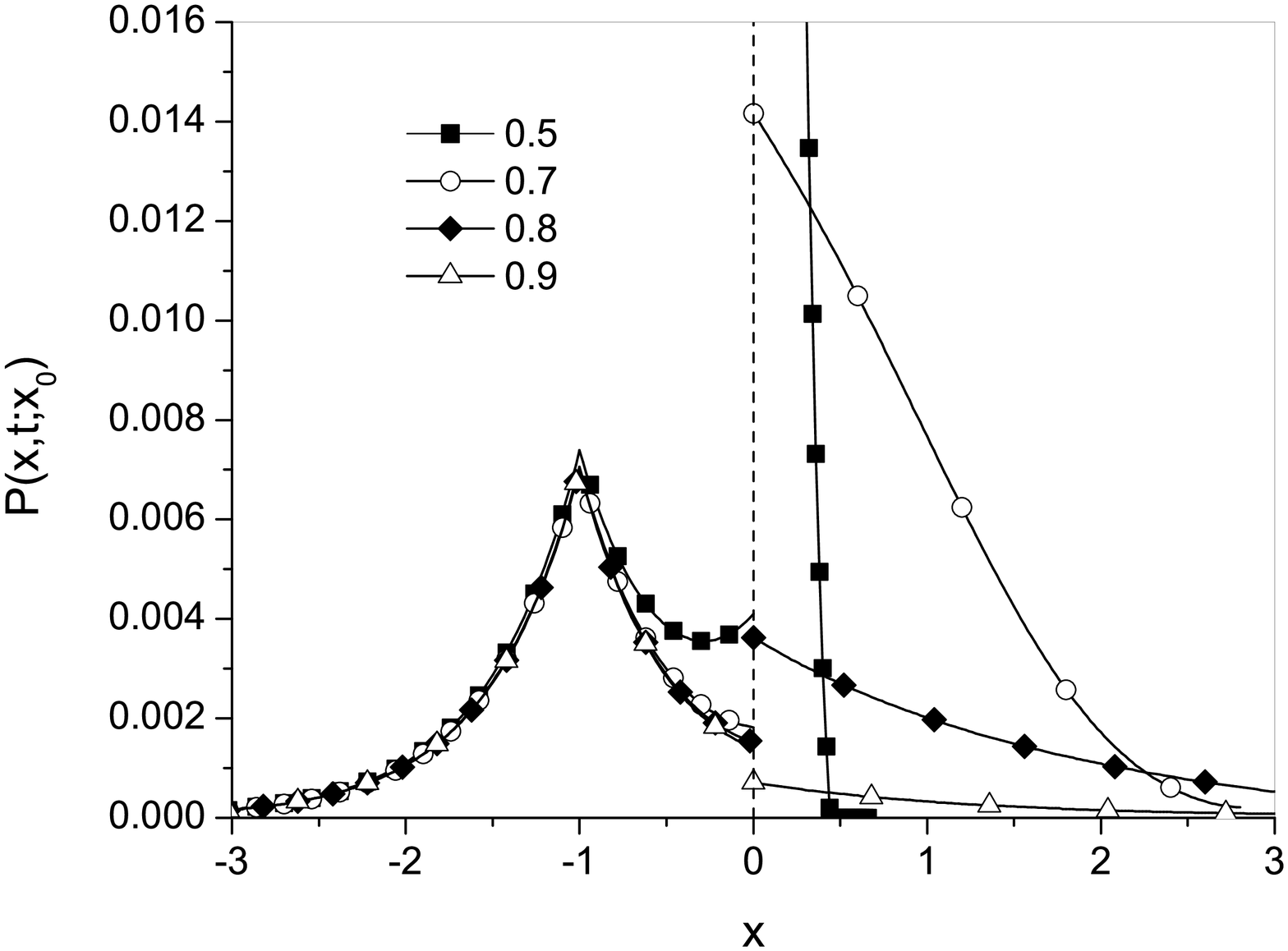}
\caption{The plots of functions (\ref{eq41}) and (\ref{eq42}) for $t=10^4$, $\alpha_1=0.9$, and for various parameter $\alpha_2$ given in the legend.}\label{Fig2}
\end{figure}

\begin{figure}[!ht]
\centering
\includegraphics[height=5.7cm]{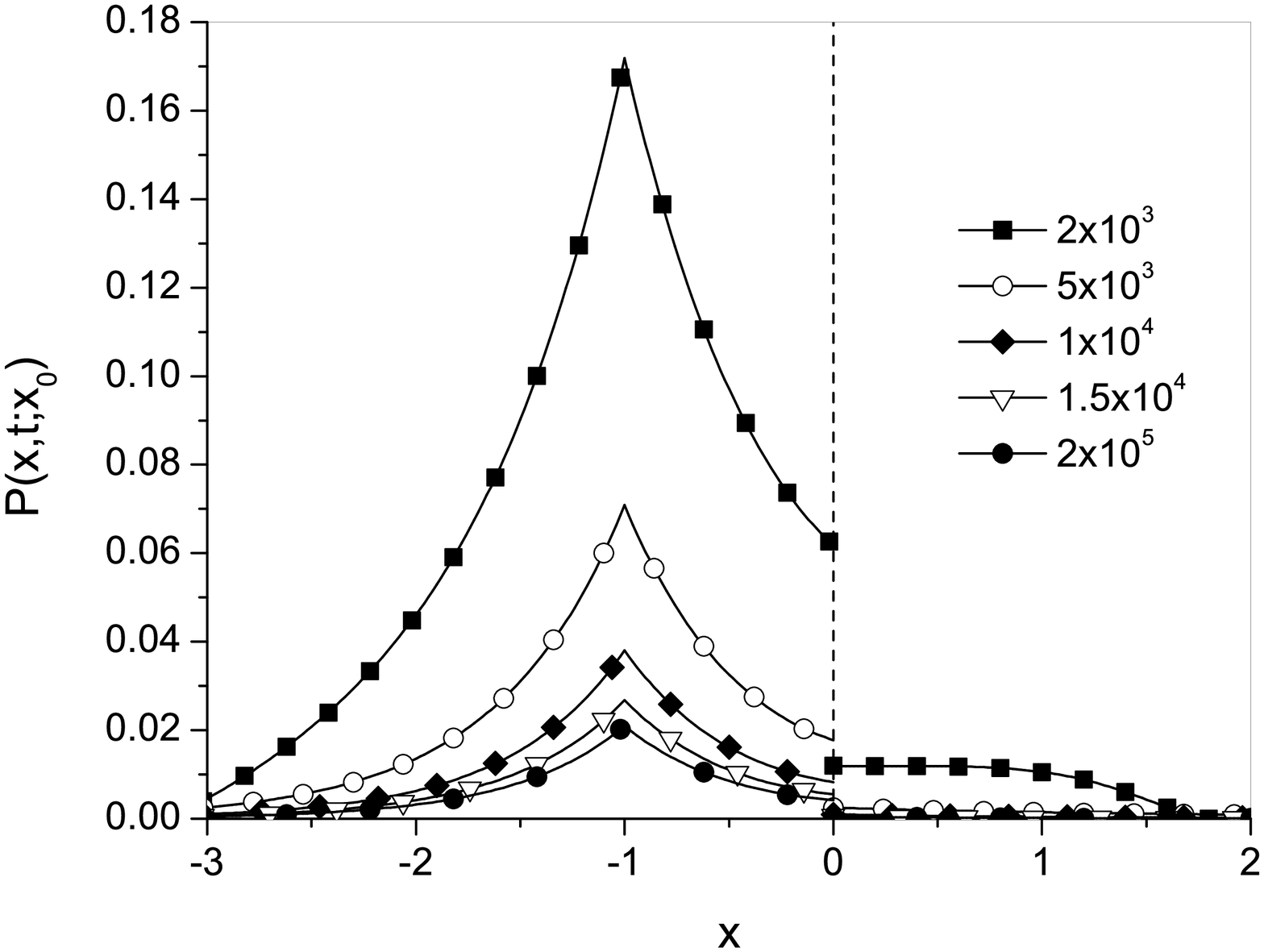}
\caption{The plots of functions (\ref{eq41}) and (\ref{eq42}) for $\alpha_1=0.8$, $\alpha_2=0.9$, and for various times given in the legend.}\label{Fig3}
\end{figure}

\begin{figure}[!ht]
\centering
\includegraphics[height=5.7cm]{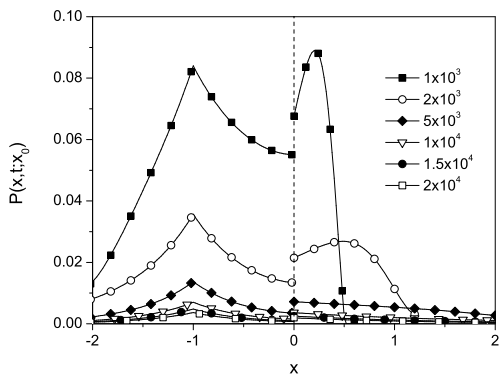}
\caption{The plots of functions (\ref{eq41}) and (\ref{eq42}) for $\alpha_1=0.9$, $\alpha_2=0.8$, and for various times given in the legend.}\label{Fig4}
\end{figure}

The plots of the functions Eqs. (\ref{eq41}) and (\ref{eq42}), presented in Figs. \ref{Fig1}--\ref{Fig4}, show that the probability of finding a particle in the regions $x<0$ and $x>0$ is mainly determined by the parameters $\alpha_1$ and $\alpha_2$. For long time, the probabilities of finding the particle in the appropriate region $W_-(t;x_0)=\int_{-\infty}^0 P_-(x,t;x_0)dx$ and $W_+(t;x_0)=\int_0^\infty P_+(x,t;x_0)dx$ read
\begin{equation}\label{eq43}
W_-(t;x_0)=\frac{1}{t^{\alpha_1}\Gamma(\alpha_1)D_1\kappa_1(\kappa_1+\kappa_2)},
\end{equation}
\begin{equation}\label{eq44}
W_+(t;x_0)=\frac{1}{t^{\alpha_2}\Gamma(\alpha_2)D_2\kappa_2(\kappa_1+\kappa_2)}.
\end{equation}
Thus, over the long time limit the probability of finding the particle in the `faster' medium (i.e. in the medium with a larger parameter $\alpha$) is negligibly small compared with the probability of finding the particle in the `slower' medium.

\section{Final remarks\label{secV}}

The main results presented in this paper are the boundary conditions, Eqs. (\ref{eq37}), (\ref{eq38}), and the Laplace transform of the Green's functions (\ref{eq28}), (\ref{eq29}). From Eqs. (\ref{eq28}) and (\ref{eq29}) we can derive Green's functions over a limit of long time for various relations between the subdiffusion parameters $\alpha_1$ and $\alpha_2$. 

Eqs. (\ref{eq28}) and (\ref{eq29}) represent the Laplace transform of the Green's functions for the case of $x_0<0$. The functions can be transformed to the case of $x_0>0$ by means of the conversions $(\alpha_1, D_1, \kappa_1)\leftrightarrow (\alpha_2, D_2, \kappa_2)$ and $(x,x_0)\leftrightarrow (-x,-x_0)$ in Eqs. (\ref{eq28}) and (\ref{eq29}). The obtained functions still fulfil the boundary conditions Eqs. (\ref{eq37}) and (\ref{eq38}). If we consider a system with many particles and if the particles move independently of each other, the concentration $C(x,t)$ can be calculated by means of the following formula $C(x,t)=\int_{-\infty}^\infty P(x,t;x_0)C(x_0,0)dx_0$. Thus, the function $C$ also fulfills the boundary conditions (\ref{eq37}) and (\ref{eq38}).

The boundary condition (\ref{eq37}) contains two fractional time derivatives. The presence of a fractional derivative proves that a process is of long memory. If $\alpha_1>\alpha_2$, the boundary condition can be transformed to the following equation $P_-(0^-,t;x_0)=(D_2/D_1)\partial^{\alpha_1-\alpha_2}P_+(0^+,t;x_0)/\partial t^{\alpha_1-\alpha_2}$. The probability of finding a particle near the border in a faster region (i.e. in the region where the parameter $\alpha$ is larger) depends on the long history of apperance of the particle on the other side of the border. Memory length depends on the difference $|\alpha_1-\alpha_2|$ (see Eq. (\ref{eq3})). The long memory effect is created only by the difference between the subdiffusion parameters in both media. If further obstacle, such as a partially permeable wall, will be located at the border beetwen media, an additional memory effect will be created \cite{tk1,kl1}.

To obtain the Green's functions in a system in which subdiffusion--reaction process occurs we apply the model of particle's random walk in a system with both discrete time and spatial variables. However, the equations describing diffusion-reaction in discrete system and in continuous system have different interpretations. A model based on discrete equations describes a process in which each jump of particle has the same length, and the absorption can only take place, with some probability, just prior to the next jump. The diffusion--reaction equation can also be obtained from the continuous time random walk model within the mean--field approximation \cite{ah}. The interpretations of both models are similar when we suppose that the position $m$ in the discrete system represents the interval $(\epsilon m-\epsilon/2,\epsilon m+\epsilon/2)$ in the system with continuous space variable and the process is considered over very long time \cite{kl}. If $\epsilon$ is assumed to be not too small, then fluctuations of particles' concentration can be neglected as in the case of the mean field approximation. However, regardless of the interpretation of both models, discrete model can be treated as a useful tool to determine the Green's functions and the boundary conditions at the border between media. The reason of such statement is that the discrete model leads to Green's functions that are solutions to the subdiffusion--reaction equations (\ref{eq1}) and (\ref{eq2}). In addition, the discrete model has a simple physical interpretation, which gives credence to the boundary conditions derived by means of this model.

As it is shown in \cite{kl}, subdiffusion with reaction $A+B\rightarrow B$ is described by Eq. (\ref{eq4}) only when the probability of meeting of particles $A$ and $B$ just after the jump made by the particle $A$ is less than 1. When the probability of particles' meeting is equal to 1, the reaction $A+B\rightarrow B$ causes the same effect as the reaction $A\rightarrow B$ \cite{kl}. Then the subdiffusion--reaction process cannot be described by Eq. (\ref{eq4}) \cite{sokolov}. The results presented in this paper cannot be applied in this case.

\section*{Acknowledgement}

The author wolud lite to thank Dr. Katarzyna D. Lewandowska for her helpful discussions. This paper was partially supported by the Polish National Science Centre under grant No. 2014/13/D/ST2/03608.

\end{document}